\documentstyle[11pt,moriond,epsfig]{article}

\def\be{\begin{equation}}
\def\ee{\end{equation}}
\def\bea{\begin{eqnarray}}
\def\eea{\end{eqnarray}}

\begin{document}
\vspace*{4cm}
\title{ON THE DETECTABILITY OF RELIC (SQUEEZED) GRAVITATIONAL WAVES}

\author{L. P. GRISHCHUK}

\address{Department of Physics and Astronomy, Cardiff University, 
Cardiff CF2 3YB, United Kingdom \\ and \\ Sternberg Astronomical
Institute, Moscow University, Moscow 119899, Russia}

\maketitle\abstracts{The expected amplitudes and spectral 
slopes of relic gravitational waves, 
plus their specific correlation properties associated with the phenomenon 
of squeezing,  may allow the registration of relic (squeezed) gravitational 
waves by the first generation of sensitive gravity-wave detectors.}  

Relic gravitational waves should have been produced by strong 
variable gravitational field of the very early Universe which 
parametrically (superadiabatically) amplified the 
zero-point quantum oscillations of the gravitational waves [1]. 
The generating mechanism is universal and relies only on the validity
of the general relativity and basic principles of quantum field 
theory. The initial vacuum quantum state of each pair of waves 
with oppositely directed momenta has evolved into a highly correlated 
multiparticle state known as the two-mode squeezed vacuum quantum 
state [2]. (For a recent review of squeezed states see [3]). 
The phenomenon of squeezing manifests itself now in a specific 
standing-wave pattern and periodic correlation functions of the
generated field [4]. It is remarkable that this statistical 
signature may significantly facilitate the detection of the 
relic (squeezed) gravitational wave background. It is possible,
as we will show below, that
the appropriate data processing will allow detection of relic 
gravitational waves by first generation of the forthcoming 
sensitive instruments, such as initial laser interferometers 
in LIGO [5], VIRGO [6], GEO600~[7].
\par
We consider cosmological gravitational wave field $h_{ij}$       
defined by the expression
\begin{equation}
\label{1}
{\rm d}s^2 = a^2({\eta})[{\rm d}\eta^2 - (\delta_{ij} + h_{ij})
{\rm d}x^i{\rm d}x^j]~.
\end{equation} 
The Heisenberg operator for the quantized real field $h_{ij}$ can 
be written as
\begin{eqnarray}
\label{2}
h_{ij} (\eta ,{\bf x})
= {C\over (2\pi )^{3/2}} \int_{-\infty}^\infty d^3{\bf n}
  \sum_{s=1}^2~{\stackrel{s}{p}}_{ij} ({\bf n})
   {1\over \sqrt{2n}}
\left[ {\stackrel{s}{h}}_n (\eta ) e^{i{\bf nx}}~
                 {\stackrel{s}{c}}_{\bf n}
                +{\stackrel{s}{h}}_n^{\ast}(\eta ) e^{-i{\bf nx}}~
                 {\stackrel{s}{c}}_{\bf n}^{\dag}  \right],
\end{eqnarray}
where $C=\sqrt{16\pi}~l_{Pl}$ and the Planck length is 
$l_{Pl}=(G\hbar /c^3)^{1/2}$, the creation and annihilation operators satisfy 
${\stackrel{s}{c}}_{\bf n}|0\rangle =0$, 
$[{\stackrel{s'}{c}}_{\bf n},~{\stackrel{s}{c}}_{{\bf m}}^{\dag}]=
\delta_{ss'}\delta^3({\bf n}-{\bf m})$, and the wave number $n$ is
related to the wave vector ${\bf n}$ by 
$n = (\delta_{ij}n^in^j)^{1/2}$. The numerical value of the constant $C$ is
determined by the quantum normalization, that is, by the requirement that, 
initially, each mode of the field has had the energy of only 
of a ``half of the quantum". 
The ${\stackrel{s}{p}}_{ij}({\bf n})$ $(s = 1, 2)$ 
are two polarisation tensors. 
The functions ${\stackrel{s}{h}}_n(\eta )$ are governed by the linearised
version of the Einstein equations. For every wave number $n$  
and each polarisation component $s$,
the functions ${\stackrel{s}{h}}_n(\eta )$ have the form 
${\stackrel{s}{h}}_n(\eta ) = 
[{\stackrel{s}{u}}_n(\eta ) + {\stackrel{s}{v}}_n^{\ast} (\eta )]/a(\eta)$,  
where ${\stackrel{s}{u}}_n(\eta )$ and ${\stackrel{s}{v}}_n(\eta )$
are expressed in terms of the three
real functions (for each s): $r_n$ - squeeze parameter, 
$\phi_n$ - squeeze angle, $\theta_n$ - rotation angle,   
\[   
u_n = e^{i{\theta}_n} \cosh~{r}_n, \qquad
v_n = e^{-i({\theta}_n - 2{\phi}_n )} \sinh~{r}_n.
\]
The functions 
$r_n(\eta)$, $\phi_n(\eta)$, $\theta_n(\eta)$ obey 
the dynamical equations [4]: 
\begin{equation}
\label{3}
r_n^{\prime} = \frac{a^{\prime}}{a} \cos{2{\phi}_n}, \quad
\phi_n^{\prime} = -n - \frac{a^{\prime}}{a} \sin{2{\phi}_n}\coth~2{r}_n, \quad 
\theta_n^{\prime} = -n - \frac{a^{\prime}}{a} \sin{2{\phi}_n}\tanh~{r}_n,
\end{equation}
where $^{\prime} = {d}/{d\eta}$, and the evolution begins 
from $r_n = 0$, which
characterizes the initial vacuum state. These equations can be transformed
into a more familiar form of the second-order 
differential equation for the function 
${\stackrel{s}{\mu}}_n(\eta ) \equiv 
{\stackrel{s}{u}}_n(\eta ) + {\stackrel{s}{v}}_n^{\ast} (\eta )$ [1]:
\begin{equation}
\mu_{n}^{\prime\prime} + \mu_{n} \left[n^2 - 
\frac{a^{\prime\prime}}{a}\right] = 0. 
\end{equation}
Dynamical equations and their solutions are identical for both polarisation 
components. 
\par
The present day values of $r_n$ and $\phi_n$ 
are essentially all we need to calculate. The mean number of particles
in a two-mode squeezed state is $2\sinh^2{r_n}$ (for each $s$). This number 
determines the 
mean square amplitude of the gravitational wave field. The time behaviour of
the squeeze angle $\phi_n$ determines the time dependence of the correlation 
functions of the field. The amplification (that is, the growth of $r_n$) 
governed 
by (3) is different for different wave numbers $n$. Therefore, 
the present day results depend on the present day frequency $\nu$
($\nu = {cn}/{2 \pi a}$) measured in $Hz$.  
\par 
In the long-wavelength regime, that is,  
during the interval of time
when the wavelength $\lambda(\eta) = 2 \pi a/n$ is longer than the Hubble 
radius $l(\eta) = a^2/a^{\prime}$,
the squeeze parameter $r_n(\eta)$ grows with time according to   
$r_n(\eta) \approx \log {[a(\eta)/{a_*}]}$, 
where $a_*$ is the value of $a(\eta)$ when the long-wavelength regime, 
for a given $n$, begins. The final amount of $r_n$ is
$r_n \approx \log {[a_{**}/a_{*}]}$, 
where $a_{**}$ is the value of $a(\eta)$ when the long-wavelength regime and 
amplification come to the end. (During the long-wavelength regime, the 
dominant solution for the function
${\stackrel{s}{h}}_n(\eta )$ is approximately constant, 
instead of the adiabatic decay 
${\stackrel{s}{h}}_n(\eta ) \sim \frac{e^{-in\eta}}{a(\eta)}$ which 
takes place in the 
short-wavelength regime.) After the end of amplification, the accumulated
(and typically large) squeeze parameter $r_n$ stays approximately constant, 
and the functions $\phi_n(\eta)$,
$\theta_n(\eta)$ are $\phi_n = -n(\eta + \eta_n)$, $\theta_n = \phi_n$.  
(For more detail see [8] and references therein.) 
The constant $\eta_n$ is determined by the moment of time when the 
long-wavelength regime, for a given $n$, terminates. This constant varies 
from one wave number $n$ to another, 
but does not change too much at the intervals $\Delta n \approx n$.
This constant will eventually survive as the (not very important) additive
constant $t_{\nu}$ in phases of the presently existing
oscillating field $\cos[2 \pi \nu(t - t_{\nu})]$, where $t$ is the total
elapsed time (billions of years since the end of the amplifying regime for
the $mHz - kHz$ frequencies of the current experimental interest), 
and therefore $\nu (t - t_{\nu}) >> 1$ for these frequencies. 
\par 
The numerical results depend on the concrete temporal behaviour of the 
gravitational pump field represented by the cosmological scale 
factor $a(\eta)$. We know that the present
matter-dominated stage $a(\eta) \propto \eta^2$ was preceeded by the 
radiation-dominated stage $a(\eta) \propto \eta$. Both these functions
are power-law functions in terms of $\eta$ time.  
The function $a(\eta)$ describing the initial stage of expansion
of the very early Universe (before
the era of primordial nucleosynthesis) is not known. It is convenient to
parameterize the $a(\eta)$ at the initial stage also by power-law functions 
of $\eta$. It is known [1] that power-law functions $a(\eta)$   
produce gravitational waves with power-law spectra in terms of $\nu$, and, 
vice versa, every piece of the generated spectrum which can be approximated
by a power-law function in terms of $\nu$ has been generated by a piece of 
evolution which can be approximated by a corresponding power-law 
function $a(\eta)$. 
\par
Concretely, we take $a_i(\eta)$ at the 
initial stage of expansion as $a_i(\eta) = l_o|\eta|^{1 + \beta}$ 
where $\eta$ time grows from $- \infty$. Expansion, that is  
increase of $a(\eta)$, at this
interval of evolution requires the parameter $\beta$ to be $\beta < -1$. This
parameter will eventually determine the spectral slope of the generated 
spectrum.
The parameter $l_o$ has the dimensionality of length and is associated with
the Hubble radius (in general, time-dependent) at the initial stage of 
expansion. In the special case $\beta = -2$ one encounters a portion of 
the De Sitter (inflation) evolution, whereby the Hubble radius remains 
strictly constant and equal to $l_o$. The ratio $l_{Pl}/l_o$ 
characterizes the ``strength" of the pump
field and will eventually determine the mean square amplitudes of the 
generated field. 
From $\eta = \eta_1$, $\eta_1 < 0$, 
the initial stage of expansion is followed by the radiation-dominated stage
$a_e(\eta) = l_oa_e(\eta - \eta_e)$ and then, from $\eta = \eta_2$,
by the matter-dominated stage 
$a_m(\eta) = l_oa_m(\eta - \eta_m)^2$. 
The constants $a_e$, $a_m$, $\eta_e$, $\eta_m$            
are expressed in terms of the fundamental parameters $l_o$, $\beta$ through
the continuous joining of $a(\eta)$ and $a^{\prime}(\eta)$ at $\eta_1$,
$\eta_2$.
The present era is defined by the observationally 
known value of the Hubble radius 
$l_H = c/H \approx 2 \times 10^{28}~{\rm cm}$.  
We denote this time by $\eta_R$ and choose $\eta_R - \eta_m = 1$,
so that $a(\eta_R) = 2l_H$. The ratio $a(\eta_R)/a(\eta_2) = z$ is believed
to be around $z = 10^4$. The wave numbers
$n_H (n_H = 4 \pi)$, $n_m (n_m = \sqrt z n_H)$, $n_c$
denote the waves which are leaving the long-wavelength 
regime at, correspondingly, $\eta_R$, $\eta_2$, $\eta_1$. 
The shorter waves, with $n > n_c$, have never been
in the amplifying long-wavelength regime.  
(The present day frequency $\nu_c$, corresponding to the wave number $n_c$,
is around $10^{10}~ Hz$.) 
\par
The most complete description, allowed by quantum mechanics, of the
generated field is provided by the evolved operator 
$h_{ij} (\eta ,{\bf x})$, eq. (2). The field is stochastic in the sense 
that it is characterised by the quantum mechanical mean values, variances,
various correlation functions, etc..  
The mean value of the field $h_{ij}$ is zero,                       
$\langle 0|h_{ij}(\eta, {\bf x})|0\rangle = 0$, at every moment of time and
in each spatial point.
The variance 
$\langle 0|h_{ij}(\eta, {\bf x})h^{ij}(\eta, {\bf x})|0\rangle ~\equiv~
\langle h^2 \rangle$ 
is not zero, and it determines the mean square amplitude of the generated
field - the quantity of interest for the experiment. Taking the product
of two expressions (2), one can show that   	 
\[
\langle h^2 \rangle = \int_0^\infty h^2(n, \eta) \frac{{\rm d}n}{n}, 
\] 
where, for the present era, 
\begin{equation}
\label{5}
h(n) \approx  A \left(n\over n_H\right)^{\beta+2}~~,~~~~~  n\leq n_H, 
\end{equation}
\begin{equation}
\label{6}
h(n,\eta) \approx A \cos\phi_n (\eta) \left(n\over n_H\right)^\beta~~, 
~~~~~ n_H \leq n\leq n_m	, 
\end{equation}
\begin{equation}
\label{7}
h(n,\eta) \approx A \cos\phi_n (\eta) \left(n\over n_H\right)^{\beta+1}  
\left(n_H\over n_m \right)~~,~~~~~  n_m \leq n\leq n_c, 
\end{equation}
and
\[
A = \frac{l_{Pl}}{l_o}\frac{8\sqrt{\pi}2^{\beta+2}}{|1+\beta|^{\beta+1}}~.
\]
The periodic time-dependent structure $\cos \phi_n (\eta)$ is a consequence
of squeezing, and it will be analyzed below. 
\par 
So far, the parameters $l_o$ and $\beta$ were quite arbitrary.
The available information on the microwave background 
anisotropies [9, 10] allows us to obtain some information 
about $A$ and $\beta$.
The quadrupole anisotropy produced by the spectrum (5) - (7) is
mainly accounted for by the wave numbers near $n_H$. Thus, 
the numerical value of the quadrupole anisotropy produced by relic 
gravitational waves is approximately equal to $A$. 
Since (according to [11]) the quadrupole contribution of relic 
gravitational waves is not smaller than that produced by primordial 
density perturbations,
this gives us $A\approx 10^{-5}$. We are not certain whether
a significant part of the quadrupole signal is indeed provided by relic 
gravitational waves,
but we can assume this. The evaluation of the spectral 
index ${\rm n}$ of the primordial perturbations resulted 
in ${\rm n} = 1.2 \pm 0.3$ [10] or even in a significantly higher 
value ${\rm n} = 1.84 \pm 0.29$ [12] (see also [13], where one of the best
fits corresponds to ${\rm n} = 1.4$). The most recent analysis [14] of
the observational data 
favors ${\rm n} = 1.2$ and the quadrupole contribution of gravitational 
waves twice as large as the contribution of density perturbations. 
We interpret these evaluations 
as an indication that the true value of ${\rm n}$ lies somewhere in the 
interval ${\rm n} = 1.2 \sim 1.4$ (hopefully, the planned new 
observational missions will determine this
index more accurately). Since the primordial (before further processing 
in short-wavelength regimes) spectral index ${\rm n}$ is related with
$\beta$ by ${\rm n} = 2\beta+5$, this gives us the 
parameter $\beta$ in the interval $\beta = -1.9 \sim -1.8$. The derived
parameters $A$, $\beta$ define the numerical level of $h(n, \eta)$ in 
the frequency
domain accessible to laser interferometers (eq. (7)) and make the 
gravitatinal wave signal measurable, as we will discuss shortly.   
\par
One can often hear that ``inflation predicts" a negligibly small 
contribution of gravitational waves, and, as a result, hopeless 
prospects for direct detection of relic gravitational waves. This is a
damaging statement, and it requires special attention. The origin of this
``conventional wisdom" is the (incorrect) prediction of inflationary 
theorists for the amplitudes of density perturbations, which, in
addition to gravitational waves, can also be produced, under some 
extra assumptions, by the same amplifying mechanism.
The central inflationary formula for the generated matter density 
variation (or curvature perturbation) is     
\begin{equation}
\frac{\delta \rho}{\rho}~~ \sim~~ \frac {H^2}{\dot\phi}~~ 
\sim~~ \frac {V^{3/2}}{V'}~,
\end{equation}
where $V(\phi)$ is the scalar field potential, and the right hand side 
is supposed
to be evaluated at the time of entering the long-wavelength regime by a 
considered mode. (For a recent summary of inflation
see, for example, [15].) Imagine that the potential $V$ has an inflection
point where $V' = 0$ ($\dot\phi = 0$). Then the predicted amplitude of
the generated mode, which happened to enter the long-wavelength regime
at that moment of time, is infinite. Since for the generation 
of a spectral interval  
with the Harrison-Zeldovich slope ${\rm n} = 1$ one needs
the denominator in the r.h.s. of (8) to be very close to zero during 
some interval of time,
all the amplitudes in this interval of spectrum are predicted 
by inflationists to be arbitrarily
large. Then, according to this logic, the relative contribution
of gravitational waves becomes negligibly small and formally zero.     
\par
There is absolutely no physical reason for this divergent result 
in a space-time (gravitational pump field) with finite and 
relatively small curvature, 
which sets the value of the nominator in (8). This result is also in
conflict with the finite ``temperature" [16] of the De Sitter space-time
determined by its constant Hubble parameter $H$.     
Whatever ``particles" are being produced, they are supposed to have finite
(and small) energy in every frequency interval. However, according to (8),
even a short interval of the De Sitter evolution is capable of generating
an arbitrarily large amount of ``particles" - density (curvature) perturbations.
It was shown [11] that formula (8) does not follow from proper evolution
and quantum normalization of density perturbations. Density perturbations
are not simply a scalar test-field, which is  
normally being considered. Density perturbations necessarily involve metric 
perturbations and do not exist without them. At every interval of evolution
with the power-law scale factor $a(\eta)$, the combined degree of freedom
satisfies exactly the same equation as eq. (4) for gravitational waves.
The associated metric (curvature) perturbation $h$ remains constant 
and small during all the long-wavelength regime, similarly to 
gravitational waves. And every interval of the power-law evolution with
parameter $\beta$ generates an interval  
of spectrum with the primordial spectral index    
${\rm n} = 2\beta+5$. In the correct version of (8), the right hand side
must be multiplied by the dimensionless factor $\sqrt{\frac{-\dot H}{H^2}}$.
This factor cancels out the zero in the denominator and makes the amplitudes
of density perturbations finite and of the same order of magnitude as 
the amplitudes of gravitational waves.
\par
The real situation on the sky looks even more interesting.   
It is important to recall [17, 11] that a confirmation of the primordial 
${\rm n} > 1$ ($\beta > -2$), even at a short spectral interval, will mean 
that the very early Universe was not driven by a scalar field - the 
cornerstone of inflationary considerations - because 
the ${\rm n} > 1$ ($\beta > -2$) 
requires the effective equation of state at the initial stage to be 
$\epsilon + p < 0$, but this cannot be accomodated by any scalar field with
whichever scalar field potential. 
\par
We switch now from cosmology to experimental predictions
in terms of laboratory frequencies $\nu$ and intervals of time $t$     
$(c{\rm d}t = a(\eta_R){\rm d}\eta)$. Formula (7) translates into
\begin{equation}
\label{9}
h(\nu,t) \approx 
10^{-7}\cos[2\pi\nu(t - t_{\nu})]\left(\frac{\nu}{\nu_H}\right)^{\beta+1},
\end{equation} 
where $\nu_H = 10^{-18} Hz$, and $t_{\nu}$ is a deterministic (not random) 
function of frequency which does not vary significantly 
on the intervals $\Delta\nu \approx \nu$. We take $\nu = 10^2 Hz$ as 
the representative frequency for the 
ground-based laser interferometers. The expected sensitivity of the 
initial instruments at $\nu = 10^2 Hz$ is
$h_{ex} = 10^{-21}$ or better. The theoretical 
prediction at this
frequency, following from (9), is $h_{th} = 10^{-23}$ for $\beta = -1.8$, 
and $h_{th} = 10^{-25}$ for $\beta = -1.9$. Therefore, the gap between
the signal and noise levels is from 2 to 4 orders of magnitude. This gap
should be covered by a sufficiently long observation time $\tau$.
The duration
$\tau$ depends on whether the signal has any temporal 
signature known in advance, or not.  
\par
It appears that the periodic structure (9) should survive in 
the instrumental window of sensitivity from $\nu_1$ (minimal frequency) 
to $\nu_2$ (maximal frequency). The mean square value of the field in this 
window is 
\begin{equation}
\label{10}
\int_{\nu_1}^{\nu_2}h^2(\nu,t)\frac{{\rm d}\nu}{\nu} =
10^{-14}\frac{1}{{\nu_H}^{2\beta+2}}\int_{\nu_1}^{\nu_2}\cos^2[2\pi\nu(t -
t_{\nu})]\nu^{2\beta+1}{\rm d}\nu~.
\end{equation}
Because of the strong dependence of the integrand on frequency, 
$\nu^{-2.6}$ or $\nu^{-2.8}$, the integral (10) is determined by 
its lower limit. This gives
\begin{equation}
\label{11}
\int_{\nu_1}^{\nu_2}h^2(\nu,t)\frac{{\rm d}\nu}{\nu} \approx
10^{-14}\left(\frac{\nu_1}{{\nu_H}}\right)^{2\beta+2}\cos^2[2\pi\nu_1(t -
t_1)]~.
\end{equation}
The explicit time dependence of the variance 
of the field, or, in other words, the explicit time dependence of 
the (zero-lag) temporal correlation function of the field, demonstrates 
that we are dealing with a non-stationary process (a consequence
of squeezing and severe reduction of the phase uncertainty). Apparently,
the search through the data should be based on the periodic 
structure at $\nu = \nu_1$.
\par
The response of an instrument to the incoming radiation is 	
$s(t) = F_{ij}h^{ij}$ where $F_{ij}$ depends on the position and 
orientation of the
instrument. The cross correlation of responses from two
instruments $\langle 0|s_1(t)s_2(t)|0\rangle$ will involve the overlap 
reduction
function [18 - 21], which we assume to be not much smaller than 1 [20]. 
The essential
part of the cross correlation will be determined by an expression of the
same form as (11).          
\par
The signal to noise ratio $S/N$ in the measurement of the amplitude 
of a signal with no specific known features increases as
$(\tau \nu_1)^{1/4}$. If 
the signal has known features exploited by the matched filtering technique, 
the $S/N$ increases as
$(\tau \nu_1)^{1/2}$. 
The guaranteed law $(\tau \nu_1)^{1/4}$ 
requires a reasonably short time $\tau = 10^6~{\rm sec}$ to improve the
$S/N$ by two orders of magnitude and to reach the level of 
the predicted signal with the extreme spectral index $\beta = -1.8$.
If the law $(\tau \nu_1)^{1/2}$ can be implemented,  
the same observation time $\tau = 10^6~{\rm sec}$ will allow the registration 
of the signal with the conservative spectral index $\beta = -1.9$. 
Even an intermediate law between
$(\tau \nu_1)^{1/4}$ and $(\tau \nu_1)^{1/2}$ 
may turn out to be sufficient. For the network of ground-based 
interferometers the expected $\nu_1$ is around
$30 Hz$, but we have used $\nu_1 = 10^2 Hz$ for a conservative 
estimate of $\tau$.
If the matched filtering technique can indeed be used, it can prove 
sufficient to have data from a single interferometer. 
\par
For the frequency
intervals covered by space intereferometers, solid-state detectors, 
and electromagnetic detectors (see also [22]), the
expected results follow from the same formula (9) and have been briefly
discussed elsewhere [17].    
\par
In conclusion,
the detection of relic (squeezed) gravitational waves may be awaiting 
only the first generation of sensitive instruments and an appropriate 
data processing strategy.   
\par
Useful discussions with S. Dhurandhar, B. Sathyaprakash, and B. Allen are
appreciated.
\par
-----------------------

\end{document}